\documentclass[pra,aps,eqsecnum,amsmath,amssymb,showpacs,showkeys]{revtex4}
\usepackage{xtheorem}
\newcommand{\am}{\mathsf{A}}

\newcommand{\ax}{\mathsf{X}}
\newcommand{\ay}{\mathsf{Y}}
\newcommand{\az}{\mathsf{Z}}
\newcommand{\tm}{\mathsf{T}}
\newcommand{\wm}{\mathsf{W}}
\newcommand{\pen}{\openone}
\newcommand{\hh}{\mathcal{H}}

\newcommand{\ip}{\mathsf{\Pi}}
\newcommand{\U}{\mathsf{U}}
\newcommand{\mem}{\mathsf{E}}

\newcommand{\tr}{{\rm{tr}}}
\newcommand{\rd}{{\rm{D}}}
\newcommand{\rs}{{\rm{S}}}
\newcommand{\ri}{{\rm{I}}}
\newcommand{\mic}{{\mathcal{I}}^R}
\newcommand{\mec}{{\mathcal{E}}^Q}
\newcommand{\me}{{\mathcal{E}}}
\newcommand{\fem}{{\mathcal{F}}}
\newcommand{\gem}{{\mathcal{G}}}
\newcommand{\re}{{\mathcal{R}}}
\newcommand{\pq}{\mathbf{p}}
\newcommand{\qr}{\mathbf{r}}
\newcommand{\qw}{\mathbf{w}}
\newcommand{\kr}{{\rm{ker}}}
\newcommand{\prq}{|\Psi^{RQ}\rangle}
\newcommand{\brq}{\langle\Psi^{RQ}|}
\newcommand{\hg}{\rho^{Q}}
\newcommand{\rrg}{\rho^{R'Q'}}
\newcommand{\kd}{\mathcal{D}}
\newcommand{\fo}{F_{\Omega}}
\newcommand{\wom}{\widetilde{\Omega}}
\newtheorem{plain}{Thm}{Theorem}[section]
{Lemma}
{Definition}
{Remark}

\begin{document}

\preprint{}

\title{{\bf Fano type quantum inequalities in terms of $q$-entropies}}

\author{Alexey E. Rastegin}
 \affiliation{Department of Theoretical Physics, Irkutsk State University,
Gagarin Bv. 20, Irkutsk 664003, Russia}
 \email{rast@api.isu.ru}

\begin{abstract}
Generalizations of the quantum Fano inequality are considered. The
notion of $q$-entropy exchange is introduced. This quantity is
concave in each of its two arguments. For $q\geq0$, the inequality
of Fano type with $q$-entropic functionals is established. The
notion of coherent information and the perfect reversibility of a
quantum operation are discussed in the context of $q$-entropies.
By the monotonicity property, the lower bound of Pinsker type in
terms of the trace norm distance is obtained for the Tsallis
relative $q$-entropy of order $q=1/2$. For $0\leq{q}\leq2$, Fano
type quantum inequalities with freely variable parameters are
obtained.
\end{abstract}



\maketitle

\section{Introduction}

In both the classical and quantum information theory, the Fano
inequality is one of the key tools. It is essential to prove the
converse to Shannon's second theorem \cite{CT91}. The quantum Fano
inequality is needed for complete proof of the quantum data
processing inequality \cite{nielsen}. Some generalizations of the
Shannon entropy have found use in various topics. One of
frequently used entropic measures was proposed by R\'{e}nyi
\cite{renyi}. Fano type inequalities in terms of R\'{e}nyi's
entropy are important in the context of classification problems
\cite{EP04}. Another variant of one-parametric extension was
introduced in classical information theory by Havrda and
Charv\'{a}t \cite{HC67} and in statistical physics by Tsallis
\cite{tsallis}. The Tsallis entropy was found to be very
significant in numerous topics of physics and other sciences
\cite{gmt}. In particular, Tsallis relative-entropy minimization
can be applied to statistical inference problems \cite{ct98,dmb}.
For $q>1$, a $q$-parametric extension of the classical Fano inequality was
given in \cite{sf06}. The entropic uncertainty principle has
been expressed in terms of both the R\'{e}nyi
\cite{zpv08,rast102} and Tsallis entropies
\cite{majer01,rast104}.

Blahut showed that the standard Fano inequality can be derived
from the properties of the relative entropy \cite{blahut}. A
development of this idea leads to a family of Fano-like
inequalities for random variables \cite{HV94}. The author of the
paper \cite{sharma} proposed extensions of quantum Fano's
inequality on the base of monotonicity of the quantum relative
entropy. For $0\leq{q}\leq2$, the Tsallis relative entropy also
enjoys the monotonicity under the action of quantum operations.
The aim of the present work is to examine Fano type quantum
inequalities in terms of Tsallis' $q$-entropies. Inequalities of
such a kind will be obtained on the base of monotonicity as well
as in another way. We also discuss a connection between the
monotonicity and lower bounds on the relative $q$-entropy. A
Pinsker type lower bound is deduced for $q=1/2$. The paper is
organized as follows. In Section \ref{tsat}, the definitions and
preliminary results are presented. A generalization of the quantum
Fano inequality in terms of Tsallis' entropies is obtained in
Section \ref{fiex}.  Lower bounds on the relative $q$-entropy are
considered in Section \ref{nopin}. In Section \ref{finer}, a
family of Fano type quantum inequalities is obtained on the base
of monotonicity property. Section \ref{concls} concludes the paper
with a summary of results.

\section{Definitions and notation}\label{tsat}

First, we recall the definitions of used entropic measures. For
real $q\geq0$ and $q\neq1$, we define the non-extensive
$q$-entropy of probability vector $\pq=(p_1,\ldots,p_n)$ by
\cite{tsallis}
\begin{equation}
S_{q}(\pq)\triangleq(1-q)^{-1} \left(\sum\nolimits_{i=1}^{n} p_i^{q}
- 1 \right)=\sum\nolimits_{i=1}^{n} \eta_q(p_i)
\ ,  \label{tsaent}
\end{equation}
where $\eta_q(x)=\left(x^q-x\right)/(1-q)$ for brevity. This can
be recast as $S_{q}(\pq)=-\sum_i p_i^{q}\ln_{q}p_i$ in terms of
the $q$-logarithm $\ln_{q}x\equiv\left(x^{1-q}-1\right)/(1-q)$,
defined for $q\geq0$, $q\neq1$ and $x>0$. The quantity
(\ref{tsaent}) will be referred to as ''Tsallis $q$-entropy'',
though it was previously discussed by Havrda and Charv\'{a}t
\cite{HC67}. In the limit $q\to1$, $\ln_{q}x\to\ln{x}$ and the
quantity (\ref{tsaent}) recovers the Shannon entropy. For any
$p\in[0;1]$, the binary Tsallis entropy is defined as
$H_q(p)=\eta_q(p)+\eta_q(1-p)$. The entropy (\ref{tsaent}) reaches
the maximal value $\ln_{q}n$ with the uniform distribution $p_i=1/n$.
For normalized density operator $\rho$ on $d$-dimensional Hilbert
space, the Tsallis $q$-entropy is defined as
\begin{equation}
\rs_q(\rho)\triangleq(1-q)^{-1}\bigl(\tr(\rho^q)-1 \bigr)=
\tr\bigl(\eta_q(\rho)\bigr)
\ . \label{tsaeq}
\end{equation}
The maximal value $\ln_{q}d$ is reached for maximally mixed state
$\pen/d$. The limit $q\to1$ leads to the von Neumann entropy
$\rs_1(\rho)=-\tr(\rho\ln\rho)$. Its general properties are
summarized in \cite{wehrl}.

In the classical regime, the relative $q$-entropy was defined as
\cite{borland}
\begin{equation}
D_q(\pq||\qr)\triangleq-\sum\nolimits_{i}p_i\ln_q(r_i/p_i)=
(1-q)^{-1}\left(1-\sum\nolimits_{i}p_i^qr_i^{1-q} \right)
\ . \label{srtdef}
\end{equation}
For basic properties of this measure, see Refs.
\cite{borland,fky04}. In particular, the relative entropy
$D_{q}(\pq||\qr)$ is monotone for all $q\geq0$ \cite{fky04}.
Namely, if $\tm=[[t_{ij}]]$ denotes the transition probability
matrix, obeying $\sum_i t_{ij}=1$ for all $j$, then
\begin{equation}
D_{q}(\tm\pq||\tm\qr)\leq D_{q}(\pq||\qr)
{\ }\qquad (0\leq q)
\ , \label{monocl}
\end{equation}
where probability vectors are put as columns. This fact easily follows from the generalized log-sum inequality
derived in \cite{borland}. In the binary case, we will write
\begin{equation}
\kd_q(u,v)\equiv D_q\bigl(\{u,1-u\}\big|\big|\{v,1-v\}\bigr)
{\ }\qquad (u,v\in[0;1])
\ . \label{binq}
\end{equation}
For $0\leq{q}<1$, a quantum extension seems to be obvious. If
$\rho$ and $\sigma$ are normalized density operators then
\cite{fky04,abe03}
\begin{equation}
\rd_{q}(\rho||\sigma)\triangleq (1-q)^{-1}\left(1-\tr(\rho^{q}\sigma^{1-q})\right)
\ . \label{reqdef}
\end{equation}
When $q>1$, the case of singular $\sigma$ should be taken into
account. The expression (\ref{reqdef}) can be adopted for
$\kr(\sigma)\subset\kr(\rho)$, otherwise
$\rd_{q}(\rho||\sigma)=+\infty$. For $0\leq
q\leq2$, the quantum relative $q$-entropy enjoys the monotonicity
under trace-preserving quantum operations. The formalism of
quantum operations provides a unified treatment of possible state
change in quantum theory \cite{nielsen}. Let $\hh$ and $\hh'$ be
finite-dimensional Hilbert spaces, and let operators $\mem_{\mu}$
map $\hh$ to $\hh'$. Any trace-preserving quantum operation $\me$
is represented as linear map \cite{nielsen}
\begin{equation}
\rho\mapsto\me(\rho)=\sum\nolimits_{m} \mem_{m}
{\,}\rho{\,} \mem_{m}^{\dagger}
\ , \label{operd}
\end{equation}
given that $\tr\bigl(\me(\rho)\bigr)=1$ for all normalized inputs
$\rho$. The last condition is equivalent to
$\sum_{m}\mem_{m}^{\dagger}\mem_{m}=\pen$, where $\pen$ is the
identity operator on $\hh$. The map (\ref{operd}) must be
completely positive as well \cite{nielsen}. The monotonicity of quantum
relative $q$-entropy implies that for any trace-preserving $\me$,
\begin{equation}
\rd_{q}\bigl(\me(\rho)||\me(\sigma)\bigr)\leq\rd_{q}(\rho||\sigma)
{\ }\qquad (0\leq q\leq2)
\ . \label{nonotd}
\end{equation}
This inequality can be obtained by applying of Lieb's concavity
theorem for $0\leq{q}<1$ and Ando's convexity theorem
for $1<q\leq2$ (for a review of this issue, see \cite{JR10}). It also follows from the general
results of the papers \cite{hmp10,naresh09}, since the function $x\mapsto{x}^q$
is matrix concave for $0\leq{q}\leq1$ and matrix convex for $1\leq
{q}\leq2$ (see, e.g., chapter V of Ref. \cite{bhatia}).

The quantum Fano inequality imposes an upper bound on the entropy
exchange in terms of the entanglement fidelity \cite{bsch96}. Here
we deal with the two quantum systems, reference system $R$ and
principal system $Q$. The initial state $\hg$ of system $Q$ is
mapped into $\mec(\hg)$. To monitor the entanglement transmission,
we consider a purification $\prq\in\hh_R\otimes\hh_Q$ which is
transformed into the final state of the joint system $RQ$ given by
\begin{equation}
\rrg=\mic\otimes\mec\bigl(\prq\brq\bigr)
\ . \label{strqe}
\end{equation}
The system $R$ itself is not altered, i.e.
$\tr_Q\bigl(\rrg\bigr)=\tr_Q\prq\brq{\,}$. The entanglement fidelity
is defined as
\begin{equation}
F\left(\hg,\mec\right)=\brq\rrg\prq
\ . \label{efid}
\end{equation}
In Ref. \cite{bsch96}, Schumacher defined the entropy exchange as
$\rs_1\left(\hg,\mec\right)=\rs_1(R',Q')=-\tr\bigl(\rrg\ln\rrg\bigr)$.
Imaging an environment $E$, we can reexpress the quantum operation
$\mec$ as
\begin{equation}
\mec(\hg)=\tr_E\left(\U_{QE}(\hg\otimes|e_0\rangle\langle e_0|)\U_{QE}^{\dagger}\right)
\ . \label{quopc}
\end{equation}
Since the final state
$(\pen_R\otimes\U_{QE})\prq\otimes|e_0\rangle$ of the triple
system $RQE$ is obviously pure, the final density operators $\rrg$
of $RQ$ and $\rho^{E'}$ of $E$ are both the partial traces of the
same one-rank projector. So these final density operators have the
same non-zero eigenvalues, whence the entropy exchange is equal to
$S_1(E')=-\tr\bigl(\rho^{E'}\ln\rho^{E'}\bigr)$ \cite{nielsen}.
Due to a similar observation, both the entanglement fidelity and
entropy exchange are not dependent on the choice of initial
purification $\prq$ \cite{bsch96}. We are now ready to quantify
the entanglement transmission by other entropic functionals.

\begin{Def}
For $q\geq0$ and $q\neq1$, the $q$-entropy exchange is defined by
\begin{equation}
\rs_{q}\left(\hg,\mec\right)\triangleq\rs_q(R',Q')=\tr\bigl(\eta_q(\rrg)\bigr)
\ . \label{sqedf}
\end{equation}
\end{Def}

As non-zero eigenvalues of the operators $\rrg$ and $\rho^{E'}$ are the
same, we have
$\rs_{q}\left(\hg,\mec\right)=\rs_q(E')=\tr\bigl(\eta_q(\rho^{E'})\bigr)$.
So this quantity characterizes an amount of $q$-entropy introduced
by the operation $\mec$ into an initially pure environment $E$.
Because the state of $E$ after the action of $\mec$ is
\cite{nielsen}
\begin{equation}
\rho^{E'}=\sum\nolimits_{mn} w_{mn} |e_{m}\rangle\langle e_{n}|
\ , \label{reaf}
\end{equation}
where
$w_{mn}=\tr\left(\mem_{m}{\,}\hg\mem_{n}^{\dagger}\right)$
are entries of the matrix $\wm$, we have
$\rs_{q}\left(\hg,\mec\right)=\tr\bigl(\eta_q(\wm)\bigr)$. In a similar
manner, the entanglement fidelity can be expressed as \cite{bsch96}
\begin{equation}
F\left(\hg,\mec\right)=\sum\nolimits_{m}\left|\tr(\hg\mem_{m})\right|^2
\ . \label{redf}
\end{equation}
The last formulae for the entropy exchange and the entanglement
fidelity are rather useful for explicit calculations. Since the
main definitions are already given, we may simplify the notation
to $\rs_q(\rho,\me)$ and $F(\rho,\me)$. That is, we will omit the
label ''$Q$'' whenever density matrices and quantum operations are
related to the principal system $Q$ solely.

The quantum $q$-entropy is concave for all $q\geq0$ \cite{raggio}.
Because the matrix $\wm$ is linear in $\rho$, the $q$-entropy
exchange $\rs_q(\rho,\me)$ is therefore concave in the first
argument $\rho$. Moreover, it is concave in the second argument,
i.e.
\begin{equation}
\theta{\,}\rs_q(\rho,\gem)+(1-\theta){\,}\rs_q(\rho,\fem)
\leq\rs_q\bigl(\rho,\theta\gem+(1-\theta)\fem\bigr)
\label{qopc}
\end{equation}
for any trace-preserving $\gem$, $\fem$ and all $\theta\in[0;1]$.
Indeed, we have
$\theta{\,}\rrg_{\gem}+(1-\theta){\,}\rrg_{\fem}=\rrg_{\me}$ for 
quantum operation $\me=\theta\gem+(1-\theta)\fem$. This claim
follows from the linearity of each quantum operation and the
representation (\ref{strqe}). So the concavity of the $q$-entropy
(\ref{tsaeq}) provides (\ref{qopc}).

\section{Quantum Fano inequality for the $q$-entropy exchange}\label{fiex}

In this section, we obtain an upper bound on the $q$-entropy
exchange in terms of the entanglement fidelity. The method of
derivation is very direct in character and similar to the
well-known proof of the standard Fano inequality. Before obtaining
the main result, we briefly recall one auxiliary statement. Let
$x\mapsto{f}(x)$ be a concave function of real scalar. Then for arbitrary
Hermitian operator $\ax$ and arbitrary normalized state
$|\psi\rangle$, there holds
\begin{equation}
\langle\psi|f(\ax)|\psi\rangle
\leq f\bigl(\langle\psi|\ax|\psi\rangle\bigr)
\ . \label{haps}
\end{equation}
To prove the claim, we take the spectral decomposition $\ax=\sum_j
x_j|x_j\rangle\langle x_j|$, whence $f(\ax)=\sum_j
f(x_j)|x_j\rangle\langle x_j|$ and
\begin{equation}
\langle\psi|f(\ax)|\psi\rangle=\sum\nolimits_j|c_j|^2 f(x_j)
\ . \label{psac}
\end{equation}
Here the numbers $c_j=\langle x_j|\psi\rangle$ are related to the
expansion $|\psi\rangle=\sum_j c_j|x_j\rangle$ and satisfy
$\sum_j|c_j|^2=1$. By Jensen's inequality for the concave function
$f$, we have
\begin{equation}
\sum\nolimits_j|c_j|^2 f(x_j)\leq f\left(\sum\nolimits_j|c_j|^2
x_j\right)=f\bigl(\langle\psi|\ax|\psi\rangle\bigr) \ .
\label{jens}
\end{equation}
Incidentally, we can observe that the functional
$\Phi(\ax)=\tr\bigl(f(\ax)\bigr)$ is concave as well, i.e.
\begin{equation}
\theta{\,}\Phi(\ay)+(1-\theta){\,}\Phi(\az)\leq\Phi\bigl(\theta\ay+(1-\theta)\az)
\label{picon}
\end{equation}
for Hermitian $\ay$, $\az$ and all $\theta\in[0;1]$. If the
$|x_j\rangle$'s are eigenstates of $\ax=\theta\ay+(1-\theta)\az$,
then we actually get
\begin{align}
\Phi(\ax)=\sum\nolimits_j   f\bigl(\langle x_j|\ax|x_j\rangle\bigr)&=
\sum\nolimits_jf\Bigl(\theta\langle x_j|\ay|x_j\rangle+(1-\theta)\langle x_j|\az|x_j\rangle\Bigr)
\nonumber\\
& \geq\sum\nolimits_j\theta
f\bigl(\langle x_j|\ay|x_j\rangle\bigr)+\sum\nolimits_j(1-\theta)f\bigl(\langle x_j|\az|x_j\rangle\bigr)
\label{phix1}\\
& \geq\theta\sum\nolimits_j\langle x_j|f(\ay)|x_j\rangle+
(1-\theta)\sum\nolimits_j\langle x_j|f(\az)|x_j\rangle
\ , \label{phix2}
\end{align}
or else
$\Phi(\ax)\geq\theta{\>}\tr\bigl(f(\ay)\bigr)+(1-\theta){\>}\tr\bigl(f(\az)\bigr)$.
Here the step (\ref{phix1}) follows from the concavity of the
function $f(x)$, the step (\ref{phix2}) follows from (\ref{haps}).
Since the function $\eta_q(x)$ is concave for $q\geq0$, the above
reasons show the concavity of the $q$-entropy. The desired upper
bound on the $q$-entropy exchange is posed as follows.

\begin{Thm}\label{teom1}
For $q\geq0$, the $q$-entropy exchange is bounded from above as
\begin{equation}
\rs_q(\rho,\me)\leq H_q\bigl(F(\rho,\me)\bigr)+\bigl(1-F(\rho,\me)\bigr)^q{\,}\ln_q(d^2-1)
\ . \label{esfin}
\end{equation}
\end{Thm}

{\bf Proof.} Let $\{|i\rangle\}$ be an orthonormal basis for the
system $RQ$ such that $|1\rangle=\prq$. We will use (\ref{haps}),
since the function $\eta_q(x)$ is concave for all $q\geq0$.
Introducing the operator
\begin{equation}
\ax'=\sum\nolimits_{i=1}^{d^2} |i\rangle\langle i|\rrg|i\rangle\langle i|
\ , \label{axs}
\end{equation}
the numbers $r_i=\langle i|\rrg|i\rangle$ are eigenvalues of
$\ax'$. Due to this fact and (\ref{haps}), we then obtain
\begin{equation}
\rs_q(\rho,\me)=\tr\bigl(\eta_q(\rrg)\bigr)=\sum\nolimits_{i=1}^{d^2} \langle i|\eta_q(\rrg)|i\rangle\leq
\sum\nolimits_{i=1}^{d^2} \eta_q(r_i)=\tr\bigl(\eta_q(\ax')\bigr)
\ , \label{hrhrr}
\end{equation}
where the right-hand side is the Tsallis $q$-entropy $S_q(\qr)$ of
$d^2$-dimensional probability vector
$\qr=\bigl(r_1,r_2\ldots,r_{d^2}\bigr)$. We also note that
$r_1=F(\rho,\me)$ by the choice of $|1\rangle$. Putting
$b_i=(1-r_1)^{-1}r_i$ for $2\leq i\leq d^2$, we get
\begin{align}
S_q(\qr)&=\eta_q(r_1)-\sum\nolimits_{i=2}^{d^2}(1-r_1)^q{\,}b_i^q{\,}\ln_q\bigl((1-r_1)b_i\bigr)
\nonumber\\
&=\eta_q(r_1)- (1-r_1)^q\sum\nolimits_{i=2}^{d^2}b_i^q\left(b_i^{1-q}\ln_q(1-r_1)+\ln_q b_i\right)
\nonumber\\
&=\eta_q(r_1)+\eta_q(1-r_1)-(1-r_1)^q\sum\nolimits_{i=2}^{d^2}b_i^q\ln_q b_i=H_q(r_1)+(1-r_1)^qS_q({\mathbf{b}})
\ , \label{sqab}
\end{align}
where we used the identity $\ln_q(xy)=y^{1-q}\ln_q x+\ln_q y$ and
$\sum_{2\leq{i}\leq{d^2}}b_i=1$. The right-hand side of
(\ref{sqab}) does not exceed $H_q(r_1)+(1-r_1)^q\ln_q(d^2-1)$,
since the $\mathbf{b}$ is a $(d^2-1)$-dimensional probability
vector. $\blacksquare$

Note that $d^2$ is replaced by $d_R{\,}d$, when the reference
system $R$ has a Hilbert space of dimension $d_R<d$ \cite{bsch96}.
The relation (\ref{esfin}) shows that if the $q$-entropy exchange
is large then the entanglement fidelity should be small enough.
The notion of mutual information is basic in classical information
theory \cite{CT91} and also used in some scenarios of quantum
information \cite{graaf,boykin,rast103}. In other aspects, a
similar role is played by the quantum coherent information
\cite{nielsen}
\begin{equation}
\ri_1(\rho,\me)=\rs_1\bigl(\me(\rho)\bigr)-\rs_1(\rho,\me)
\ . \label{cohin}
\end{equation}
By analogy, we can define the coherent $q$-information as
\begin{equation}
\ri_q(\rho,\me)\triangleq\rs_q\bigl(\me(\rho)\bigr)-\rs_q(\rho,\me)
\ . \label{qohin}
\end{equation}
The right-hand side of (\ref{qohin}) looks similar to the
$f$-generalization of the coherent information treated in
\cite{naresh09}. But the above expression is actually not a
partial case of such generalization. We now recall that the
Tsallis $q$-entropy enjoys the subadditivity property for $q>1$,
namely
\begin{equation}
\rs_q(Q,E)\leq\rs_q(Q)+\rs_q(E)
\ . \label{qsub}
\end{equation}
The inequality has been conjectured by Raggio \cite{raggio} and
later proved by Audenaert \cite{auden07}. Raggio also conjectured
that the inequality (\ref{qsub}) is saturated if and only if
either of the systems $Q$ and $E$ is being in a pure state, and
proved this in a partial case. It seems that equality conditions
for (\ref{qsub}) are beyond the scope of the subadditivity proof
given in \cite{auden07}. Here the question to be answered is
whether the equality in (\ref{qsub}) implies that either of two
subsystems is being in a pure state. Using (\ref{qsub}), we can
derive a triangle type inequality
\begin{equation}
\left|\rs_q(Q)-\rs_q(E)\right|\leq\rs_q(Q,E)
{\ }\qquad (1<q)
\ . \label{qsub1}
\end{equation}
The proof is easy. Introducing the reference system $R$, one
purifies systems $Q$ and $E$. Due to (\ref{qsub}), we then have
\begin{equation}
\rs_q(R,Q)\leq\rs_q(R)+\rs_q(Q)
\ . \label{qsub2}
\end{equation}
When state of the triple system $RQE$ is pure,
$\rs_q(R,Q)=\rs_q(E)$ and $\rs_q(R)=\rs_q(Q,E)$. These two
equalities allows to rewrite (\ref{qsub2}) in form
\begin{equation}
\rs_q(E)-\rs_q(Q)\leq\rs_q(Q,E)
\ . \label{qsub3}
\end{equation}
By a parallel argument, we get $\rs_q(Q)-\rs_q(E)\leq\rs_q(Q,E)$,
and the last two inequalities provide (\ref{qsub1}). This
treatment allows further extension to many of the quantum unified
entropies \cite{rastjst}. These entropies were introduced and
motivated in \cite{hey06}. We can now establish an upper bound on
the coherent $q$-information.

\begin{Thm}\label{teom2}
For $q>1$, the coherent $q$-information is bounded from above by
\begin{equation}
\ri_q(\rho,\me)\leq\rs_q(\rho)
\ . \label{qqdpr}
\end{equation}
Assuming Raggio's conjecture on equality conditions in
(\ref{qsub}), the equality in (\ref{qqdpr}) implies that the
quantum operation $\me$ is perfectly reversible upon input of
$\rho$.
\end{Thm}

{\bf Proof.} Using the definition (\ref{qohin}) and the triangle
inequality, we obtain
\begin{equation}
\ri_q(\rho,\me)=\rs_q(Q')-\rs_q(E')\leq\rs_q(Q',E')
\ . \label{qses}
\end{equation}
Note that the operation $\me$ is realized by some unitary
transformation of the space $\hh_Q\otimes\hh_E$ and the initial
state of environment $E$ is pure (see the formula (\ref{quopc})).
These points imply the equality
$\rs_q(Q',E')=\rs_q(Q,E)=\rs_q(\rho)$, which together with
(\ref{qses}) provides (\ref{qqdpr}). Assume that Raggio's
conjecture on equality conditions holds. The equality in
(\ref{qqdpr}) can be rewriten as
\begin{equation}
\rs_q(R',E')=\rs_q(R')+\rs_q(E')
\label{subeq}
\end{equation}
due to $\rs_q(Q')=\rs_q(R',E')$ and $\rs_q(Q',E')=\rs_q(R')$ (the
final state of the triple system $RQE$ is pure). So either of the
systems $R$ and $E$ should be in a pure state, whence
$\rho^{R'E'}=\rho^{R'}\otimes\rho^{E'}$ (for the last claim, see
\cite{raggio}). This product structure immediately implies an existence of
the recovery operation $\re$ such that the entanglement
fidelity of combined operation $F(\rho,\re\circ\me)=1$ (for an
explicit construction of $\re$, see the proof of theorem 12.10
in \cite{nielsen}). In other words, the quantum operation $\me$ is
perfectly reversible upon input of $\rho$. $\blacksquare$

So the coherent $q$-information enjoys, in a less degree, similar
properties to the coherent information (\ref{cohin}). The standard
data processing inequality also tells that for $q=1$ the perfect
reversibility of $\me$ upon input of $\rho$ leads to the equality
in (\ref{qqdpr}). This statement is based on the quantum Fano
inequality and the strong subadditivity property \cite{nielsen}.
In the classical regime, the Tsallis entropy of order $q>1$ obeys
the strong subadditivity \cite{sf06}. However, this result cannot
be used, since  the systems $R$ and $E$ become entangled after
action of the operation $\me$. Using another definition, the
author of \cite{naresh09} has extended the data processing
inequality in complete setting to a wide class of matrix convex
functions. Note that the function $x\mapsto{x}^q$ does enjoy the
matrix convexity for $1\leq{q}\leq2$, but does not for $2<q$ (see,
e.g., exercise V.2.11 in \cite{bhatia}). On the other hand,
Theorem \ref{teom2} holds for all $1<q$. In general, possible ways
to extend the standard concept of  coherent information
deserve further investigations.

\section{Notes on Pinsker type inequalities}\label{nopin}

Lower and upper bounds on some functional allows to estimate it in
terms of other measures or parameters. When states are close to
each other in the trace norm sense, corresponding bounds
characterize continuity of a functional. Estimates of such a kind
are important due to a statistical interpretation of the trace
distance in terms of POVM measurements \cite{nielsen}. The
partitioned trace distances also enjoy this property for one-rank
POVMs \cite{rast091}. The well-known upper bound of desired type
is given by Fannes' inequality for the von Neumann entropy
\cite{fannes}. This treatment has been extended to the Tsallis
$q$-entropy \cite{yanagi,zhang} and its partial sums
\cite{rast1023}. For the standard relative entropy, lower and
upper continuity bounds are obtained in the paper \cite{AE05}. For the relative $q$-entropy,
some upper continuity bounds were given in \cite{rast1009}. The
well-known Pinsker type lower bound is expressed as
\begin{equation}
\rd_1(\rho||\sigma)\geq\frac{1}{2}{\>}\|\rho-\sigma\|_1^2
\ , \label{qpins}
\end{equation}
where the Schatten 1-norm is $\|\ax\|_1=\tr\sqrt{\ax^{\dagger}\ax}$ for any
operator $\ax$. In much more general setting, this inequality was
proved in \cite{hots61}. It is also known \cite{still90} that
\begin{equation}
\rd_p(\rho||\sigma)\geq\rd_1(\rho||\sigma)\geq\rd_q(\rho||\sigma)
\ , \label{p1qin}
\end{equation}
where $q\in[0;1)$ and $p\in(1;2]$. So the upper bounds given in
\cite{AE05} hold for the relative $q$-entropy of order
$q\in[0;1)$, the lower ones hold for the relative $q$-entropy of
order $q\in(1;2]$. Thus, we are rather interested in lower bounds for
the former and in upper bounds for the latter. Upper continuity bounds on the relative $q$-entropy of order
$q\in(1;2]$ have recently been obtained in \cite{rast1009}. Below
we will discuss lower continuity bounds that follow from the
monotonicity of the relative $q$-entropy.

\begin{Thm}\label{pinmon}
Let $\ip_{+}$ be a projector on the eigenspace corresponding to
positive eigenvalues of the difference $(\rho-\sigma)$. For
$q\in[0;2]$ and any pairs of density operators, the relative
$q$-entropy is bounded from below as
\begin{equation}
\rd_q(\rho||\sigma)\geq\kd_q(u,v)
\ , \label{piniq}
\end{equation}
where $u=\tr(\ip_{+}\rho)$ and $v=\tr(\ip_{+}\sigma)$.
\end{Thm}

{\bf Proof.} Let us write the Jordan decomposition of traceless
Hermitian operator
\begin{equation}
\rho-\sigma=\sum\nolimits_{r>0} r{\,}|r\rangle\langle r|-
\sum\nolimits_{s>0} s{\,}|s\rangle\langle s|
\ . \label{jorde}
\end{equation}
We define the two projectors $\ip_{+}=\sum_r|r\rangle\langle r|$,
$\ip_{-}=\sum_s|s\rangle\langle s|$. When the difference
$(\rho-\sigma)$ has zero eigenvalues, corresponding eigenvectors
should be included to the orthonormal sets $\{|r\rangle\}$ and
$\{|s\rangle\}$ anyhow; then $\ip_{+}+\ip_{-}=\pen$. Consider the
trace-preserving quantum operation
\begin{equation}
\fem(\rho)=\sum\nolimits_r|r\rangle\langle r|\rho|r\rangle\langle
r| + \sum\nolimits_s|s\rangle\langle s|\rho|s\rangle\langle
s|=\sum\nolimits_{r} u_r|r\rangle\langle r| +\sum\nolimits_s
u_s|s\rangle\langle s| \ , \label{qogr}
\end{equation}
where probabilities $u_r=\langle r|\rho|r\rangle$ and $u_s=\langle
s|\rho|s\rangle$. Putting $v_r=\langle r|\sigma|r\rangle$ and
$v_s=\langle s|\sigma|s\rangle$, we also write
\begin{equation}
\fem(\sigma)=\sum\nolimits_r v_r|r\rangle\langle r|
+\sum\nolimits_s v_s|s\rangle\langle s|
\ . \label{qogs}
\end{equation}
So the outputs $\fem(\rho)$ and $\fem(\sigma)$ are diagonal in the
same basis. Due to this fact and the monotonicity of quantum
relative $q$-entropy for $0\leq q\leq2$, we have
\begin{equation}
\rd_q(\rho||\sigma)\geq\rd_q\bigl(\fem(\rho)||\fem(\sigma)\bigr)
=D_q\bigl(\{u_r,u_s\}\big|\big|\{v_r,v_s\}\bigr)
\ . \label{moncor}
\end{equation}
We shall again use the monotonicity, but now in classical regime.
Let us put the 2-by-$d$ transition probability matrix
\begin{equation}
\tm=
\begin{pmatrix}
 1 & \cdots & 1 & 0 & \cdots& 0 \\
 0 & \cdots & 0 & 1 & \cdots& 1 \\
\end{pmatrix}
\ , \label{tt2}
\end{equation}
in which the units of the first row act on $r$-components, the
units of the second row act on $s$-components. This matrix maps
the distributions $\{u_r,u_s\}$ and $\{v_r,v_s\}$ to $\{u,1-u\}$
and $\{v,1-v\}$ respectively with $u=\sum_ru_r=\tr(\ip_{+}\rho)$,
$v=\sum_rv_r=\tr(\ip_{+}\sigma)$. By the monotonicity, the
right-hand side of (\ref{moncor}) is not less than $\kd_q(u,v)$.
$\blacksquare$

In general, the projector $\ip_{+}$ and the probabilities $u$, $v$
are not uniquely defined. But for any choice, we have $u-v=\sum_r
r=(1/2){\,}\|\rho-\sigma\|_1$. The next stage is to estimate
$\kd_q(u,v)$ from below in terms of the quantity $|u-v|$. Really,
we would like to find the minimum of $\kd_q(u,v)$ under the
conditions $0\leq u\leq1$, $0\leq v\leq1$ and $|u-v|=t$. For the
standard case $q=1$, this issue is well developed (see
\cite{fedotov} and references therein). A complete examination of
the problem would take us to far afield. We consider only the case
$q=1/2$, which allows simple calculations.

\begin{Lem}\label{guvt}
In the domain $\{(u,v):{\>}0\leq u\leq1,{\>}0\leq
v\leq1,{\>}|u-v|=t\in(0;1)\}$, there holds
\begin{equation}
g(u,v)=\sqrt{uv}+\sqrt{(1-u)(1-v)}\leq
\sqrt{1-t^2}
\ . \label{gfund}
\end{equation}
\end{Lem}

{\bf Proof.} The domain consists of the two segments, the segment
$u=t+v$ with $v\in[0;1-t]$ and the segment $v=t+u$ with
$u\in[0;1-t]$. Due to symmetry, we consider the former. By
differentiating with respect to $v$ under the constraint $u=t+v$,
we obtain the condition for critical points in a form
\begin{equation}
\frac{v+u}{2\sqrt{uv}}=\frac{(1-v)+(1-u)}{2\sqrt{(1-u)(1-v)}}
\ , \label{crpt}
\end{equation}
which is clearly satisfied with $u+v=1$. Combining
this with $u=t+v$ gives $u_0=(1+t)/2$, $v_0=(1-t)/2$, and
$g(u_0,v_0)=\sqrt{1-t^2}$. Some inspection shows that the above
critical point is unique on the chosen segment (the solution $u=v$ of (\ref{crpt}) holds only for $t=0$). The value
$g(u_0,v_0)$ is actually maximal, since the function is concave
and $g(t,0)=g(1,1-t)=\sqrt{1-t}$. $\blacksquare$

Using the relations $\kd_{1/2}(u,v)=2\bigl(1-g(u,v)\bigr)$, $t=(1/2)\|\rho-\sigma\|_{1}$, and
$1-\sqrt{1-t^2}\geq t^2/2$, we finally get
\begin{equation}
\rd_{1/2}(\rho||\sigma)\geq 2-2\left(1-\frac{1}{4}{\>}\|\rho-\sigma\|_1^2\right)^{1/2}
\geq\frac{1}{4}{\>}\|\rho-\sigma\|_1^2
\ . \label{pins}
\end{equation}
This is a quantum lower bound of Pinsker type on the relative
$1/2$--entropy. As expressed in terms of the trace norm distance,
it characterizes a continuity property. The lower bounds
(\ref{qpins}) and (\ref{pins}) are independent and consistent in
view of (\ref{p1qin}). In a similar manner, lower bounds of
Pinsker type could be obtained for other values from the
interval $q\in(0;1)$. By the statement of Theorem \ref{pinmon},
the problem is merely reduced to minimization of $\kd_q(u,v)$
under certain conditions. Except for the case
$q=1/2$, an answer is not so obvious. In principle, this issue might
be a subject of separate research.

\section{Fano type inequalities for $0\leq{q}\leq2$}\label{finer}

In this section, a family of Fano type bounds on the $q$-entropy
exchange will be derived from the monotonicity of relative
$q$-entropy. In the regular case $q=1$, this idea has been
developed for the classical Fano inequality \cite{HV94} as well as
for the quantum one \cite{sharma}. The basic point is to relate
the $q$-entropy exchange with the relative $q$-entropy by
\begin{equation}
\rd_q\bigl(\rrg\big|\big|\wom\bigr)=-\rs_q(R',Q')
-\tr\left(\bigl(\rrg\bigr)^q\ln_q(\wom)\right)
\ , \label{sqex}
\end{equation}
which follows from the identity
$(1-q)^{-1}\left(x-x^qy^{1-q}\right)=-\eta_q(x)-x^q\ln_qy$ and the
normalization. By $\wom$ we denote an arbitrary nonsingular
density matrix of appropriate dimensionality. The $q$-entropy
exchange is bounded from above in the following way.

\begin{Thm}\label{teom3}
Let $\wom$ be a nonsingular density matrix on the space
$\hh_R\otimes\hh_Q$, $\prq\in\hh_R\otimes\hh_Q$ a purification of the input $\rho$ of the operation $\me$, $F_e=F(\rho,\me)$, and $\fo=\brq\wom\prq$. For $0\leq
q\leq2$, there holds
\begin{equation}
\rs_q(\rho,\me)\leq -\kd_q(F_e,\fo) -\tr\left(\bigl(\rrg\bigr)^q\ln_q(\wom)\right)
\ . \label{finom}
\end{equation}
\end{Thm}

{\bf Proof.} Let $\{|i\rangle\}$ be an orthonormal basis in $\hh_R\otimes\hh_Q$ such that $|1\rangle=\prq$. We consider the
trace-preserving quantum operation $\gem$ acting as
\begin{align}
 & \gem\bigl(\rrg\bigr)=\sum\nolimits_{i=1}^{d^2} |i\rangle\langle i|\rrg|i\rangle\langle i|
\ , \label{gerh}\\
 & \gem\bigl(\wom\bigr)=\sum\nolimits_{i=1}^{d^2} |i\rangle\langle i|\wom|i\rangle\langle i|
\ . \label{geom}
\end{align}
Both the above outputs are diagonal in the basis $\{|i\rangle\}$. Combining
this fact with the monotonicity for $q\in[0;2]$, we further write
\begin{equation}
\rd_q\bigl(\rrg\big|\big|\wom\bigr)\geq
\rd_q\left(\gem\bigl(\rrg\bigr)\big|\big|\gem\bigl(\wom\bigr)\right)
=D_q(\qr||\qw)
\ , \label{qumon}
\end{equation}
where the probabilities $r_i=\langle i|\rrg|i\rangle$ and
$w_i=\langle i|\wom|i\rangle$. We now apply the
monotonicity in classical regime with the 2-by-$d^2$ transition
probability matrix such that
\begin{equation}
\tm=
\begin{pmatrix}
 1 & 0 & \cdots& 0 \\
 0 & 1 & \cdots& 1 \\
\end{pmatrix}
\ , \qquad
\tm\qr=
\begin{pmatrix}
 r_1  \\
 1-r_1 \\
\end{pmatrix}
\ , \qquad
\tm\qw=
\begin{pmatrix}
 w_1  \\
 1-w_1 \\
\end{pmatrix}
\ . \label{ttt2}
\end{equation}
By the monotonicity, we then get $D_q(\qr||\qw)\geq\kd_q(F_e,\fo)$ in view of $r_1=F_e$ and $w_1=\fo$. Combining this with
the relations (\ref{sqex}) and (\ref{qumon}) finally gives
(\ref{finom}). $\blacksquare$

In the right-hand side of (\ref{finom}), the first term is
negative, the second is positive. The relation (\ref{finom}) with
freely variable terms presents some family of upper bounds on the
$q$-entropy exchange. Choosing various forms of $\wom$, we will be
arrived at different upper bounds. A nonlinear power $q$ of the
density operator $\rrg$ is difficult to any transformations when
this operator is not given in an explicit form. Nevertheless, for
$q\geq1$ there holds
\begin{equation}
\bigl(\rrg\bigr)^q\leq\rrg
\ , \label{rrqr}
\end{equation}
since eigenvalues of a density matrix does not exceed one. We also
have ${\,}-\ln_q(\wom)\geq{\mathbf{0}}$, due to $\ln_{q}x\leq0$ for
$x\in(0;1]$. Because $\ax\leq\ay$ implies
$\tr(\ax\am)\leq\tr(\ay\am)$ for any $\am\geq{\mathbf{0}}$, the
inequalities (\ref{finom}) and (\ref{rrqr}) lead to
\begin{equation}
\rs_q(\rho,\me)\leq -\kd_q(F_e,\fo) -\tr\left(\rrg\ln_q(\wom)\right)
{\ }\qquad (1\leq q\leq2)
\ . \label{finom1}
\end{equation}
If the principal system $Q$ is initially prepared in the state
\begin{equation}
\rho=\sum\nolimits_{k=1}^{d}\lambda_k{\,}|\lambda_k\rangle\langle\lambda_k|
\ , \label{qlam}
\end{equation}
then any purification has the form
\begin{equation}
\prq=\sum\nolimits_{k=1}^{d} \sqrt{\lambda_k}{\>}|\xi_k\rangle\otimes|\lambda_k\rangle
\label{xam}
\end{equation}
with some orthonormal basis $\{|\xi_k\rangle\}$ in $\hh_R$. For a
probability distribution $\{\mu_j\}$, we can take
\begin{equation}
\wom=\sum\nolimits_{j=1}^{d} \mu_j{\,}|\xi_j\rangle\langle\xi_j|\otimes\omega=
\sum\nolimits_{jk} \mu_j\nu_k{\,}|\xi_j\nu_k\rangle\langle\xi_j\nu_k|
\ , \label{takm}
\end{equation}
where $\omega=\sum_k\nu_k{\,}|\nu_k\rangle\langle\nu_k|$ is a
density operator on $\hh_Q$. By calculations, one obtains
\begin{equation}
\fo=\sum\nolimits_{ijk} \sqrt{\lambda_i\lambda_k}{\,}\mu_j{\,}\langle\xi_i|\xi_j\rangle{\,}\langle\xi_j|\xi_k\rangle
{\,}\langle\lambda_i|\omega|\lambda_k\rangle=
\sum\nolimits_{j=1}^d \lambda_j\mu_j \langle\lambda_j|\omega|\lambda_j\rangle
\ . \label{lmik}
\end{equation}
Using the identity $\ln_q(xy)=\ln_q x+x^{1-q}\ln_q y$, we also find
\begin{equation}
\ln_q(\wom)=\sum\nolimits_{jk} \ln_q(\mu_j\nu_k)|\xi_j\nu_k\rangle\langle\xi_j\nu_k|
=\sum\nolimits_j \ln_q\mu_j{\,}|\xi_j\rangle\langle\xi_j|\otimes\pen_Q+
\sum\nolimits_j \mu_j^{1-q}{\,}|\xi_j\rangle\langle\xi_j|\otimes\ln_q(\omega)
\ . \label{lnqw}
\end{equation}
Using (\ref{strqe}) and the linearity of $\me$, we further obtain
\begin{equation}
\rrg=\sum\nolimits_{ij}\sqrt{\lambda_i\lambda_j}
{\>}|\xi_i\rangle\langle\xi_j|\otimes\me(|\lambda_i\rangle\langle\lambda_j|)
\ . \label{reqq}
\end{equation}
We also observe that
$\tr\bigl(\me(|\phi\rangle\langle\psi|)\bigr)=\langle\psi|\phi\rangle$
by the preservation of the trace. Hence the trace of the product
of (\ref{lnqw}) and (\ref{reqq}) is written as
\begin{equation}
\tr\left(\rrg\ln_q(\wom)\right)=\sum\nolimits_{j} \lambda_j{\,}\ln_q\mu_j+
\sum\nolimits_{j}\mu_j^{1-q}\lambda_j{\>}\tr\Bigl(\me(|\lambda_j\rangle\langle\lambda_j|)\ln_q(\omega)\Bigr)
\ , \label{trem}
\end{equation}
where both the $\{\mu_j\}$ and $\omega$ are still arbitrary.
Combining (\ref{lmik}) and (\ref{trem}) with (\ref{finom1}), we
obtain an upper bound of Fano type, in which both the probability
distribution $\{\mu_j\}$ and state $\omega$ are freely variable.

A next question is, whether the $q$-parametric extension of Fano inequality
(\ref{esfin}) can be derived from
(\ref{finom}). It seems that the answer is negative in general.
For $q=1$, the quantum Fano inequality is get by
$\wom=(\pen_R\otimes\pen_Q)/d^2$ (for details, see \cite{sharma}).
For $q>1$, such a choice leads to an inequality which includes the
right-hand side of (\ref{esfin}) with some additional terms.
However, these terms are not negative anywhere. We refrain from
presenting the calculations here. Moreover, any corollary of
(\ref{finom}) would be restricted to $q\in[0;2]$, whereas the
inequality (\ref{esfin}) holds for all $q\geq0$. On the other
hand, the relation (\ref{finom}) with freely variable parameters
may lead to new inequalities similar to (\ref{finom1}). The
results of this section are essentially based on the monotonicity
of the relative $q$-entropy for $0\leq q\leq2$. We finally note
that the classical Fano inequality deals with the conditional
entropy \cite{CT91}, which is not a direct classical analog of the
quantum entropy exchange. So the quantum formulation enough
differs from the classical one. In this regard, any extension of
results of the papers \cite{blahut,HV94} to generalized entropic
functionals would be interesting.

\section{Conclusions}\label{concls}

We have considered various extensions of the quantum Fano
inequality in terms of $q$-entropic measures. The notion of the
$q$-entropy exchange was introduced with some discussion of its
properties. In particular, the $q$-entropy exchange is concave in
the input density matrix as well as in the running quantum
operation. The standard quantum Fano inequality is generalized for
all $q\geq0$ in Theorem \ref{teom1}. This result is essentially
based on the properties of the function $\eta_q(x)$ and the
related functional inequality (\ref{haps}). We have also
introduced a $q$-parametric extension of the coherent information.
Due to the subadditivity for $q>1$, the triangle inequality
(\ref{qsub1}) holds. Using this result, the upper bound on the
coherent $q$-information is posed in Theorem \ref{teom2}. Assuming
Raggio's conjecture, the inequality (\ref{qqdpr}) is saturated
only if the quantum operation $\me$ is perfectly reversible upon
input of $\rho$.

We have also obtained some bounds based on the monotonicity of the
relative $q$-entropy for $0\leq q\leq2$. For all $q\in[0;2]$, a
simple lower bound on the relative $q$-entropy is established by
Theorem \ref{pinmon}. Hence a lower continuity bound of Pinsker type
has been obtained for $q=1/2$ in the result (\ref{pins}). An
extension to other values of parameter $q$ is briefly discussed. The
monotonicity property has been used for obtaining a family of Fano
type quantum inequalities on the $q$-entropy exchange. This
statement is formulated in Theorem \ref{teom3}. Except for $q=1$,
the inequality of Theorem \ref{teom1} seems to be not included in
the presented family. Nevertheless, several interesting
inequalities with freely variable parameters can be dealt. These
inequalities might be useful in specialized problems, when some
prior knowledge on both the input state and running quantum
operation is available.

\acknowledgments
The present author is grateful to anonymous referee for constructive criticism.

\end{document}